# Observation of the Molecular Assisted Recombination in a Hydrogen Plasma

Akira TONEGAWA, Hiroyuki YAZAWA, Kentarou KUMITA,
Masataka ONO, and Kazutaka KAWAMURA

*Department of Physics, School of Science, Tokai University*
*1117Kitakaname, Hiratsuka, Kanagawa 259-1292, Japan*

We have presented the experimental observation of the spatial structure of Molecular Assisted Recombination (MAR) including both dissociative recombination and mutual neutralization channels in the detached hydrogen plasma in the linear divertor plasma simulator, TPD-SheetIV. It is shown from the results of mass-analysis ($H_2^+$, $H_3^+$) that dissociative recombination is dominant in the center of the plasma over a range of low gas pressures. At the same time, it is observed that the mutual neutralization in MAR via $H^-$ ion formation, which is produced by dissociative electron attachment to $H_2(v)$, occurs in the periphery of the plasma where cold electrons (~1 eV) are found.

1. Introduction

The knowledge of atomic and molecular processes in detached divertor regions has become more important, as a dynamic gas target regime is considered to be one of the most favourable conditions for divertor plasmas of fusion reactor to control the heat load on the divertor targets [1,2]. In the detached divertor conditions with lower plasma temperature, vibrationally excited hydrogen molecules $H_2(v)$ persist in dissociation and ionization processes of the plasma volume recombination. Thus, the plasma volume recombination associated with vibrationally excited hydrogen molecules $H_2(v)$, that is, Molecular Assisted Recombination (MAR) is effective in the divertor plasma to enhance the reduction of ion particle flux [2-5]. The rate coefficient for MAR is much greater than that one for the radiative and three-body recombination, that is, Electron-Ion Recombination (EIR) at relatively high electron temperatures above 1.0 eV [6-7]. The vibrationally excited molecules $H_2(v)$ contribute to plasma volume recombination due to the following chains of reactions: (1) $H_2(v) + e => H^- + H$ (dissociated attachment) followed by $H^- + H^+ => H + H$ (mutual neutralization), and (2) $H_2(v) + H^+ => H_2^+ + H$ (ion conversion) followed by $H_2^+ + e => H + H$ or $H_2^+ + H_2 => H_3^+ + H$ (dissociative recombination). However, the role of the MAR in fusion experiments is still under discussion and various conclusions are derived from the analysis of different experiments [8-11].

In this paper, we present the experimental observation of spatial structure of MAR for hydrogen detached plasma in the linear divertor plasma simulator, TPD-SheetIV (Test Plasma produced by Directed current for Sheet plasma) [12,13]. In order to investigate the dissociative recombination and mutual neutralization in MAR, measurements of the molecular and atomic ion densities ($n_{H^+}$, $n_{H2^+}$, $n_{H3^+}$), the negative ion density of hydrogen atom ($n_{H^-}$), the electron density ($n_e$), electron temperature ($T_e$), and the heat load to the target plate (Q) were carried out in hydrogen detached plasma with hydrogen gas puff. It is also intended to show that the observed emission intensity (the Lyman-bands $B^1\Sigma_g^+ \to X^1\Sigma_g^+$) of VUV (vacuum ultraviolet) wavelength region from electronic excited hydrogen molecules $H_2(B^1\Sigma_u^+)$ by electron impact could be explained by MAR.

2. Experimental apparatus and method

The experiment was performed in the linear divertor plasma simulator TPD-SheetIV as shown in Fig.1. The hydrogen sheet plasma was produced by a modified TPD type dc discharge[13]. Ten rectangular magnetic coils formed a uniform magnetic field of 0.04 T in the experimental region. The hydrogen plasma was generated at a hydrogen gas flow of 70 sccm with a discharge current of 50 A. The neutral pressure $P_{Div}$ in the divertor test region was controlled by feeding a secondary gas from 0.1 to 20 mTorr. The change of $P_{Div}$ in the divertor test region had no effect on the plasma production in the discharge region because the pressure difference between the discharge and divertor test regions extends to 3 orders of magnitude. The electron temperature and the electron density were measured by a plane Langmuir probe. The plane Langmuir probe was located 3 cm apart from the target plate. The power on the target plate Q was measured by a calorimeter. A cylindrical probe made of tungsten ( 0.4 x 2 cm) was used to measure the spatial profiles of $H^-$ by a probe-assisted laser photodetachment method [14]. The maximum Nd-YAG laser energy at the fundamental wavelength (1064 nm) was 100 mJ. The pulse length was about 10 ns and the diameter of the beam was 8 mm. The negative ion density was determined from the photodetached electron current. The spatial profile of the negative ions was measured by moving the cylindrical probe under the laser irradiation. An "omegatron" mass-analyzer, situated behind a small hole ( 0.5 mm) of the endplate with the differential pumping system, is used for analyzing ion species[15]. This analyzer yields signals due to the cyclotron resonances of ion species. The peaks appear in a collector current of the analyzer when a frequency of applied radio frequency (RF) electric field is equal to the ion cyclotron frequencies. The molecular and atomic ion density was determined from the collector current of the mass-analyzer. The vibrationally excited hydrogen molecules $H_2(X^1\Sigma_g^+$ (v>3-7)) result from the deexcitation of electronic excited hydrogen molecules $H_2(B\Sigma_u, C\Pi_u)$ by electron impact [16]. When this process occurs in plasmas, vacuum ultraviolet lights are emitted. Thus it is an effective approach to measure the spectra of VUV light emission of $H_2(B\Sigma_u)$ in recombination plasma of MAR. The plasma through a viewing port installed at the sidewall of the divertor test region is observed by the VUV spectrometer with the differential pumping system and a CCD camera. The spectra of VUV light emission from electronic excited hydrogen molecules $H_2(B\Sigma_u)$ were detected 3 cm apart from the target plate.

## 3. Experimental results and discussion

Figure 2 shows the dependence of the heat to the target plate, Q, the hydrogen Lyman spectrum ratio, $L_\gamma/L_\alpha$, the electron density, $n_e$, the electron temperature, $T_e$, the intensity of VUV emission spectrum at the Lyman-band system $(B^1\Sigma_u^+ \rightarrow X^1\Sigma_g^+)$, and the molecular and atomic ion density at the center of sheet plasma on hydrogen gas pressure $P_{Div}$. The VUV emission intense line spectrum is observed at the Lyman-band system $(B^1\Sigma_u^+ \rightarrow X^1\Sigma_g^+)$ in hydrogen molecular bands. The Lyman-band at 125.74 nm (12,7) is identified [16]. The densities of the molecular and atomic ions $n_H^+$, $n_{H2}^+$, and $n_{H3}^+$ were determined from the collector current ($H^+$, $H_2^+$, $H_3^+$) of the "omegatron" mass-analyzer, respectively. With increasing in $P_{Div}$, the value of Q is found to decrease rapidly, until

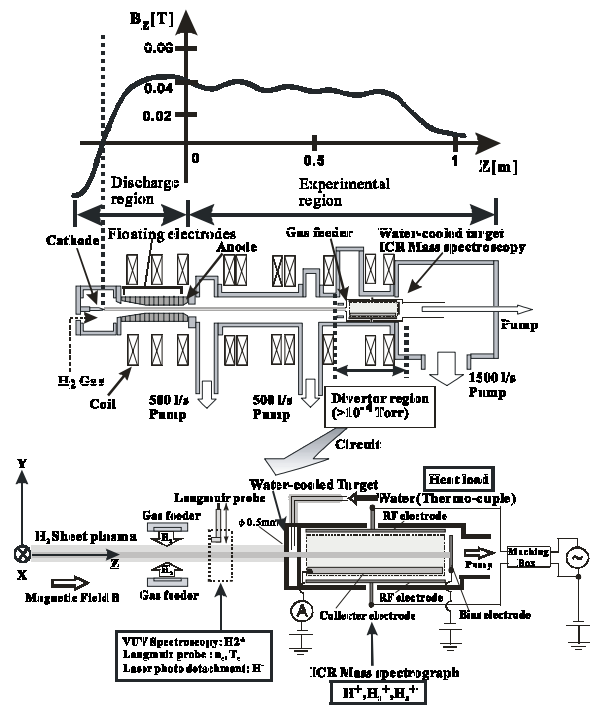

Fig.1 Schematic diagram of the magnetized plasma simulator (sheet plasma device) TPD-Sheet-IV and detection system. The profile of the axial magnetic field $B_z$ is shown in the upper part

less than 30% of the initial value at $P_{Div}$ ~ 3 mTorr, while the radiative and three-body recombination processes disappeared. The value of $n_e$ increases slightly from $4\times10^{18}$ to $5\times10^{18}$ m$^{-3}$ and $T_{ec}$ decreases rapidly from 15 to 10 eV due to ionization until $P_{Div}$ ~ 2 mTorr. At the same time, the densities of hydrogen molecular ions ($H_2^+$, $H_3^+$) and the intensities of VUV spectra at the Lyman-band system ($B^1\Sigma_u^+ \to X^1\Sigma_g^+$) of $H_2$ rapidly decrease. Above $P_{Div}$ ~ 2 mTorr in which $T_e$ is less than ~8 eV, the value of $n_e$ gradually decreases from $5\times10^{18}$ m$^{-3}$ to several $10^{16}$ m$^{-3}$. Also, we can observe a nearly constant of the intensities of VUV spectra ($B^1\Sigma_u^+ \to X^1\Sigma_g^+$) of $H_2$. Therefore, it is shown from these results that the dominant molecular process is the dissociative recombination process via $H_2^+$, $H_3^+$ in the central region of the sheet plasma over the range of low hydrogen pressure. With an increase in $P_{Div}$, the value of Q is found to decrease rapidly from 0.32 to 0.1 MW/m$^2$, a value less than 30% of the initial value in the attached plasma for $P_{Div}$ less than 3.0 mtorr, while the hydrogen Lyman spectrum ratio $L_\gamma/L_\alpha$ of VUV emission remains nearly constant. With an increase of $P_{Div}$ to ~ 4 mtorr, $n_H^-$ disappears and $L_\gamma/L_\alpha$ is observed in front of the target plate. The ratio $L_\gamma/L_\alpha$ rapidly increases above $P_{Div}$ ~ 6 mtorr. At the same time, the intensities of the hydrogen Lyman series from n=3 to 6 due to the EIR were observed in front of the target (full detachment region), that is, the radiative and three-body recombination processes have appeared. It is found that feedback control on the negative ion density is the only process allowing a factor of two reduction of the initial heat flux onto the target without strong radiation loss from EIR.

Figure 3 shows the effect of hydrogen neutral gas pressure, $P_{Div}$, on the heat to the target plate, Q, the electron density, $n_e$, the electron temperature, $T_e$, the intensity of VUV emission spectrum at the Lyman-band system ($B^1\Sigma_u^+ \to X^1\Sigma_g^+$), and the maximum value in the plasma of the hydrogen negative ion density, $n_H^-$ at a discharge current of 50 A. The electron density has a hill-shaped profile with a half width of about 5.0 mm in the sheet plasma. Also, the sheet plasma has a steep electron temperature gradient over a plasma thickness of several millimeters: a hot plasma (~15 eV) in the central region and a cold plasma (1-2 eV) in the periphery region [12-14]. The value of $n_e$ increases slightly, and $T_e$ decreases rapidly from 3 to 2 eV until $P_{Div}$ ~ 2 mtorr. Above $P_{Div}$~2 mtorr, $T_e$ falls below less than 2 eV, and $n_e$ gradually decreases. In this range of low hydrogen pressure (< 2 mTorr), vibrationally excited hydrogen molecules $H_2(X^1\sum_g^+(v>3-7))$ result from the deexcitation of electronic excited hydrogen molecules $H_2(B^1\sum_u^+)$ by the collision of $H_2$ with hot electron ($T_e$=10-15 eV) in the central region and diffuse to the periphery region of the sheet plasma. The $H^-$ ions of the periphery are localized in a periphery region of about 10 to 20 mm distance from the center in the direction of thickness of the sheet plasma. Using a small amount of secondary hydrogen gas puffing into a hydrogen plasma, $n_{H^-}$ has a maximum value of $1.0 \times 10^{17}$ m$^{-3}$ at $P_{Div}$ ~ 3.0 mtorr.

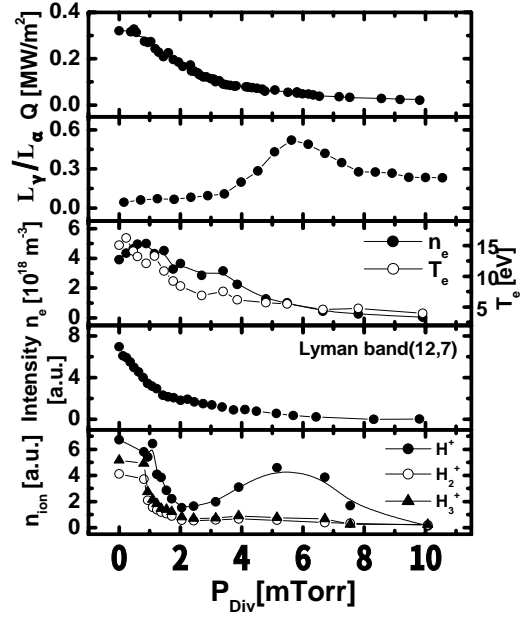

Fig.2 The dependence of Q, $L_\gamma/L_\alpha$, $n_e$, $T_e$, the intensity of VUV emission spectrum and the molecular and atomic ion density at the center of sheet plasma on hydrogen gas pressure $P_{Div}$.

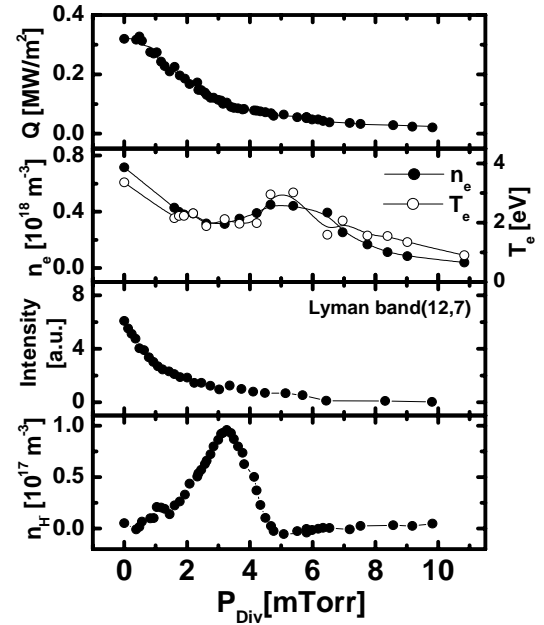

Fig.3 The effect of $P_{Div}$ on Q, $n_e$, $T_{ec}$, the intensity of VUV emission spectrum and hydrogen negative ion density, $n_H^-$ at a discharge current of 50 A.

## 4. Conclusion

The spatial profiles of Molecular Assisted Recombination (MAR) with vibrational hydrogen molecules have been observed in a detached hydrogen plasma by using a linear plasma simulator, TPD-SheetIV. From the VUV emission (Lyman-bands:$B^1\sum_u^+ \rightarrow X^1\sum_g^+$) of hydrogen molecules $H_2$, it is indicated that the production process of vibrationally excited hydrogen molecules $H_2(X^1\sum_g^+(v>3-7))$ is the deexcitation of electronic excited hydrogen molecules $H_2(B^1\sum_u^+)$, which are produced by collisions between the $H_2$ and hot electrons ($T_e$=10-15 eV) at the center of the sheet plasma. The densities of $H_2^+$ and $H_3^+$ in the central region of plasma reach a peak in the range of low hydrogen pressure (< 2 mTorr) (dissociative recombination). With increasing the hydrogen pressure from 2 to 4 mTorr, the negative hydrogen ions $H^-$, which are produced by the collision of assisted species with cold electrons ($T_e$=0.5-0.7eV) in the periphery of the plasma, recombine with hydrogen positive ion $H^+$ (mutual neutralization).

## Acknowledgments

This work is part of a program supported by the LHD Joint Project, the National Institute for Fusion Science.

## References
[1] G.Janeschitzs et al., J.Nucl.Mater. **220-222,** 73 (1995).
[2] Post D E, J.Nucl.Mater. **220-222**, 143 (1995).
[3] S.I.Krasheninnikov et al., Phys. Plasma **4,** 1637 (1997).
[4] S.I.Krasheninnikov et al., Phys. Lett. A **314,** 285 (1996).
[5] A.Yu.Pigarov et al., Phys. Lett. A **222,** 251 (1996).
[6] J.Hiskes, Rev.Rci.Instr. **63,** 2702 (1992).
[7] D. Reiter et al., J. Nucl. Mater. **241-243,** 342 (1997).
[8] N.Ohno et al., Phy.Rev.Lett. **81,** 818 (1998).
[9] A.Tonegawa et al., J. Nucl. Mater. **313-316,** 1046 (2003).
[10] P.C.Stangeby, The Plasma Boundary of Magnetic Fusion Devices, Institute of Publishig, Bristol(2000) Chap.l6.
[11] U.Fantz et al., J. Nucl. Mater. **290-293,** 367 (2001).
[12] K.Sunako et al., Nucl.Inst.&Meth. **B111,** 151 (1996).
[13] A.Tonegawa et al., J.Advanced Science **11,** 232 (1999).
[14] M.Bacal, Rev.Sci.Inst. **71**, 3981 (2000).
[15] H.Sommer et al., Phys. Rev.**82**, 697 (1951).
[16] J.M.Ajello et al., Phys.Rev. A **25**, 2485 (1982).